\documentclass[article]{aa}  

\usepackage{graphicx}
\usepackage{amsmath}
\usepackage{tikz}
\usepackage{pgf}
\usepackage{pgfplots}
\pgfplotsset{compat=newest}
\usepackage{txfonts}
\usepackage{comment}
\usepackage{booktabs}

\usepackage{xcolor}

\newcommand{\HI}{H{\small{I}}}
\defcitealias{Bianchetti2025}{B25}
\defcitealias{Donevski2023}{D23}

\usepackage{hyperref}

\begin{document} 

\title{Atomic hydrogen reservoirs in quiescent galaxies at $z=0.4$}

   \subtitle{}

   \author{A. Bianchetti\inst{1,2}
          \and
          G. Rodighiero\inst{1,2}
          \and
          D. Donevski\inst{3,4}
          \and
          F. Sinigaglia\inst{5,6}
          \and
          E. Elson\inst{7}
          \and 
          M. Vaccari\inst{8,9,10}
          \and
          A. Marasco\inst{2}
          \and
          L. Bisigello\inst{2}
          \and 
          I. Prandoni\inst{10}
          \and
          M. Baes\inst{11}
          \and 
          M. Glowacki\inst{12,8}
          \and 
          F. M. Maccagni\inst{13}
          \and 
          G. Lorenzon\inst{3}
          \and 
          I. Heywood\inst{14,15,16}
          }

   \institute{Department of Physics and Astronomy, Università degli Studi di Padova, Vicolo dell’Osservatorio 3, I-35122, Padova, Italy
        \and INAF - Osservatorio Astronomico di Padova, Vicolo dell’Osservatorio 5, I-35122, Padova, Italy
        \and National Center for Nuclear Research, Pasteura 7, 02-093 Warsaw, Poland
        \and SISSA, Via Bonomea 265, 34136 Trieste, Italy
        \and Département d’Astronomie, Université de Genève, Chemin Pegasi 51, 1290, Versoix, Switzerland
        \and Institut für Astrophysik, Universität Zürich, Winterthurerstrasse 190, CH-8057 Zürich, Switzerland
        \and Department of Physics and Astronomy, University of the Western Cape, Robert Sobukwe Rd, 7535 Bellville, Cape Town, South Africa
        \and Inter-university Institute for Data Intensive Astronomy, Department of Astronomy, University of Cape Town, 7701 Rondebosch, Cape Town, South Africa
        \and Inter-university Institute for Data Intensive Astronomy, Department of Physics and Astronomy, University of the Western Cape, 7535 Bellville, Cape Town, South Africa
        \and INAF - Istituto di Radioastronomia, via Gobetti 101, 40129 Bologna, Italy
        \and Sterrenkundig Observatorium, Universiteit Gent, Krijgslaan 281 S9, 9000 Gent, Belgium22
        \and Institute for Astronomy, University of Edinburgh, Royal Observatory, Edinburgh, EH9 3HJ, United Kingdom
        \and INAF – Osservatorio Astronomico di Cagliari, via della Scienza 5, I-09047 Selargius, CA, Italy
        \and Astrophysics, Department of Physics, University of Oxford, Keble Road, Oxford, OX1 3RH, UK
        \and Centre for Radio Astronomy Techniques and Technologies, Department of Physics and Electronics, Rhodes University, PO Box 94, Makhanda, 6140, South Africa
        \and South African Radio Astronomy Observatory, 2 Fir Street, Black River Park, Observatory, Cape Town, 7925, South Africa
             }

   \date{Received July 11, 2025; accepted September 10, 2025}

  \abstract
   {Based on local Universe observations, quiescent galaxies (QGs) host lower or negligible \HI{} compared to star-forming galaxies (SFGs), but no constraints have been derived to date at higher redshift ($z>0.1$). Understanding whether QGs can retain significant \HI{} reservoirs at higher z is crucial for refining quenching and gas accretion models and for constraining overall star formation efficiency at different epochs.
   }
   {We aim to probe \HI{} in candidate QGs at intermediate redshifts (\rm $\langle z \rangle \approx 0.36$) and to understand whether a class of QGs exists that retains consistent \HI{} reservoir, as well as which parameters (dust content, stellar mass, $D_n$4000, morphology, and environment) effectively capture \HI{}-rich QGs.}
   {We performed 21-cm spectral line stacking on MIGHTEE-HI data at $\langle$z$\rangle=0.36$, targeting two different samples of QGs, defined by means of a color-selection criterion and a spectroscopic criterion based on $D_n$4000, respectively. We also performed stacking on subsamples of the spectroscopically selected quiescent sample to investigate the correlation between the \HI{} content and other galaxy properties.
   }
   {We find that QGs with an IR counterpart (i.e., dusty galaxies) host a substantial \HI{} content, on average only $40\%$ lower than that of SFGs. In contrast, color-selected QGs still retain \HI{}, but at levels lower than those of SFGs by a factor of $\sim$3. Among dusty objects, we find that morphology has a mild impact on the atomic gas content, with spirals hosting approximately $15-30\%$ more \HI{} than spheroids. Environmental effects are also present: galaxies in low-density regions are richer in \HI{} than those in high-density regions, by approximately $30\%$ for spirals and $60\%$ for spheroids. We suggest that, in general, \HI{} content is influenced by several factors, including slow quenching mechanisms and interstellar medium (ISM) enrichment processes. Also, QGs --- and especially dusty systems --- seem to yield \HI{} more consistently than in the local Universe.}
   {}


   \maketitle
%
\section{Introduction} \label{sec:intro}
Although not directly fueling star formation, neutral atomic hydrogen (\HI{}) is a key phase of the interstellar medium (ISM), particularly at high redshift, becoming dominant  at $z\gtrapprox2$ \citep{Peroux2020, Walter2020}. It offers insight into quenching and gas retention and rejuvenation. It can be detected through the 21-cm line arising from the hyperfine structure of the atom. The intrinsic faintness of this line makes direct detection very challenging: current record holders for high-redshift direct 21-cm detection are a set of six galaxies detected by FAST at z$\approx$0.4 \citep{Xi2024}. For this reason, the \HI{} community employed indirect statistical methods to measure average \HI{} properties at higher distances. In particular, spectral stacking has yielded statistically robust results at redshifts up to $<\approx0.4$ \citep[e.g., see][]{Lah2007, Delhaize2013, Bera2019, Chowdhury2020, Guo2020, Guo21, Chowdhury2022a, Sinigaglia2022, Bera2023, Rhee2023, Sinigaglia2024, Bianchetti2025}, where only a limited number of direct detections are currently available. Pioneering uGMRT (upgraded Giant Meter Radio Telescope, see \citealt{Gupta2017}) studies pushed the boundaries of stacking studies to $z\approx1$ \citep{Chowdhury2021, Chowdhury2022b, Chowdhury2022d}.

Although traditionally regarded as gas-poor, recent works provide evidence of cold molecular gas in quiescent galaxies \citep[QGs;][]{Combes2007, Davis2013, Gobat2018}. In particular, analysis of the ATLAS-3D sample \citep{Cappellari2011} reveal signficant amounts of atomic gas in early-type galaxies \citep[ETGs;][]{Serra2012, Maccagni2017, yildiz20}. \citet{Young2014} further showed that even these ETGs may have quenched from blue progenitors while retaining a fraction of their cold gas budget. Subsequent studies using the Sloan Digital Sky Survey \citep[SDSS;][]{Abazajian2009}, the Arecibo Legacy Fast ALFA survey \citep[ALFALFA;][]{Giovanelli2005}, the GALEX Arecibo SDSS Survey \citep[GASS;][]{Catinella2010}, and the COLD GASS survey \citep{Saintonge2011} show evidence that the vast majority of massive quiescent disk galaxies contain abundant \HI{} reservoirs \citep{Zhang2019}.

In contrast, other studies find a lack of \HI{} in QGs. In the local Universe, \citet{Cortese2020} find little to no \HI{} down to $\rm \log{M_{HI}} \approx 8$ in quiescent spirals from the extragalactic GASS survey \citep[xGASS;][]{Catinella2018}, claiming that reduced star formation correlates with depletion of \HI{} reservoirs. \citet{Gereb2016, Gereb2018} report only isolated cases of \HI{}-rich QGs, labeling them as \HI{}-excess galaxies. Building on this, \citet{Saintonge2022} note that the existence of galaxies below the main sequence (MS) with relevant \HI{} reservoirs does not contradict the general trend of \HI{}-poor QGs, as these galaxies have low \HI{} fractions. At slightly higher redshift ($\rm 0.04<z<0.09$), stacking color-selected QGs in the Deep Investigation of Neutral Gas Origins (DINGO; \citealt{Meyer2009}) survey reveals no \HI{} down to $\rm \log{M_{HI}} \approx 9$ \citep{Rhee2023}. These contradictory results make the presence of atomic gas in QGs a debated issue. 

The \HI{} content of QGs beyond the local Universe remains relatively unexplored and could provide a better understanding of star formation efficiency throughout cosmic time.
In this paper, we stack samples of candidate QGs in the COSMOS field (from \citet{Bianchetti2025} and \citet{Donevski2023}, hereafter \citetalias{Bianchetti2025} and \citetalias{Donevski2023}) at a mean redshift $\rm z\approx 0.37$. We use data from the MeerKAT International Giga-Hertz Tiered Extragalactic Exploration survey \citep[MIGHTEE;][]{Jarvis2016, Heywood2024}) obtained with the MeerKAT radio interferometer \citep{Jonas2016} to study the presumed \HI{} content in the two samples of quenched galaxies. The paper is organized as follows. In Sect.~\ref{sec:sample_selection} we briefly summarize the specifics of the MIGHTEE-HI radio survey used for the stacking analysis, present the samples used, and briefly discuss the steps of the selection process. Sect.~\ref{sec:results} outlines the main results of stacking the samples in different slices of phase space, analyzing the impact of dust presence, stellar mass, $D_n$4000, morphology, and environment, and interpreting our findings in the framework of modern results on galaxy evolution. Sect.~\ref{sec:conclusions} summarizes the paper. We assume a flat ($\Omega_k$=0) Lambda-CDM cosmology, employing cosmological parameters from \citet{Planck2020}, i.e., $H_0=67.4~{\rm km~s}^{-1}~ {\rm Mpc^{-1}}$, $\Omega_m=0.315$, and $\Omega_{\Lambda}=0.685$, and an initial mass function from \citet{Chabrier2003}. 

\section{Data and sample} 
\label{sec:sample_selection}
In this work, we stack spectra extracted from MIGHTEE-HI DR1 \citep{Heywood2024}, a survey consisting of 15 deep MeerKAT pointings (94.2 h total) within the COSMOS field. The survey covers two separate frequency ranges (960-1150 MHz and 1290-1520 MHz), where the former is used in this paper, targeting the \HI{} 21 cm line in the redshift range $0.22<z<0.5$. Comprehensive details on data reduction and data products are provided by \citet{Heywood2024}.

We used two different catalogs of candidate QGs, both covering the COSMOS field, as input for the stacking pipeline. The first catalog defines quiescence using a double rest-frame color criterion ($NUV-r$ vs $r-J$) and is the same catalog used in \citetalias{Bianchetti2025}. We refer to that paper for a full description. The second catalog is inherited by \citetalias{Donevski2023} and originally comprises 4362 galaxies at $0.01<z<0.7$ with spectroscopic data from the hCOSMOS survey\citep{Damjanov2018}. We provide here a brief summary of the catalog construction, but we refer to \citetalias{Donevski2023} for a full description. Here, the quiescent nature of the selected objects is identified using a combination of criteria. First, a threshold is set on high 4000 \AA-break strength ($D_n4000 > 1.5$), ensuring robust separation between star-forming galaxies (SFGs) and candidate QGs (i.e., \citealt{Balogh1999}, \citealt{damjanov25}). Furthermore, objects in the catalog meet at least one additional criterion, either (i) low specific star formation rate (sSFR), placing objects more than five times below the \cite{speagle14} main-sequence, or (ii) UVJ color criteria for QGs \citep{muzzin13}. We supplemented this catalog with IR counterparts from the Herschel Extragalactic Legacy Project \citep[HELP;][]{Shirley2019} to assess the dust properties of the sample. \citetalias{Donevski2023} define a candidate QG as dusty if it is detected (S/N>3) in at least four bands at $8 \le \lambda \le 500 \mu m$, with a reference 1$\sigma$ sensitivity of $\rm \approx 1 mJy$ at $\rm \lambda=250\mu m$. Galaxies that do not meet this criterion are considered ordinary, dust-poor QGs.

\begin{figure}
    \begin{center}
        \includegraphics[width=\columnwidth]{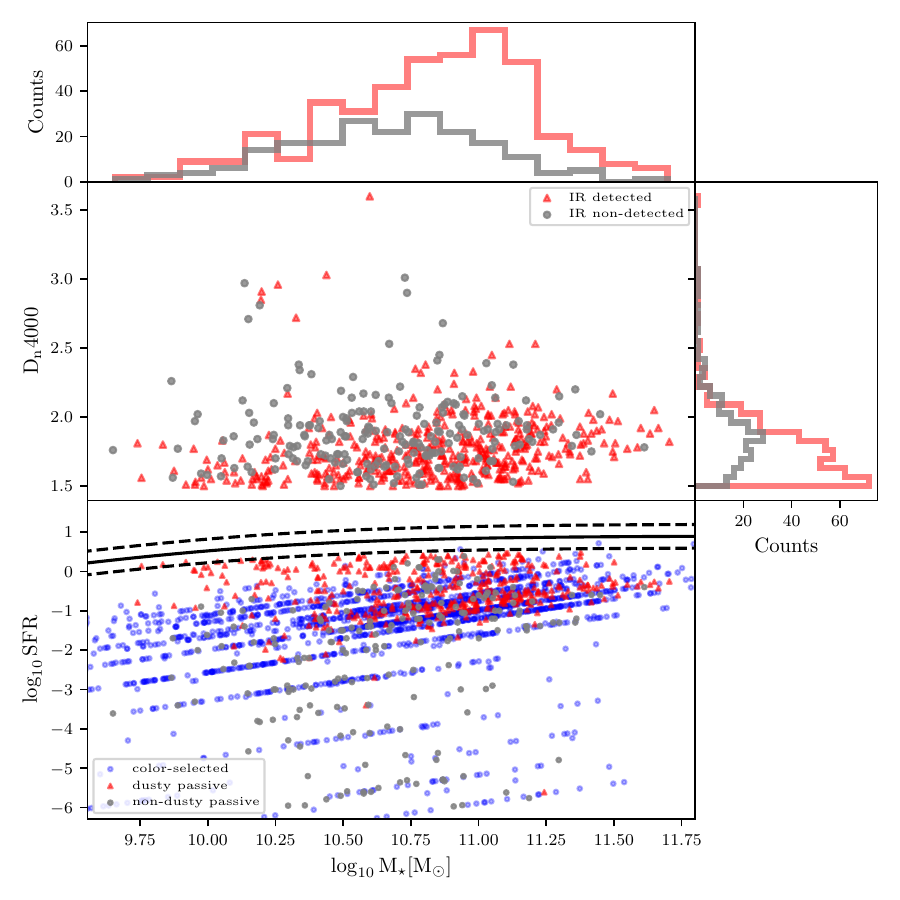}
    \end{center}
    \caption{Two-dimensional distributions of galaxy properties from three investigated samples. Upper panel: Scatter plot of the spectroscopic sample from \citetalias{Donevski2023} in the $D_n$4000-$\log{M_{\star}}$ plane. Red triangles indicate galaxies with an IR counterpart, while gray circles mark galaxies with no IR detection. Histograms above and right of the scatter plot show the distribution of these parameters for the two infrared populations, using the same color coding.
    Lower panel: Distribution of galaxies from the spectroscopic sample of \citetalias{Donevski2023}, plotted in comparison to the MS galaxies shown in the left panel. Red points represent galaxies with an IR counterpart, while gray points represent galaxies without an IR counterpart. The sample is compared with an MS model by \citet{Popesso2023} (black solid line), while the dashed black line indicates a 0.3 dex scatter around the MS.
        }
    \label{fig:spec_sample}
\end{figure}

We show the properties of the investigated sample in Fig.~\ref{fig:spec_sample}. The upper panel shows the $\log{M_{\star}}-Dn_{4000}$ parameter space in Fig.~\ref{fig:spec_sample}, where red points indicate galaxies with an IR counterpart and gray points indicate those with no detected IR counterpart, i.e., below the detection limits previously specified. Histograms at the side of the panel show the distributions of $D_n$4000 and $M_{\star}$ for the dusty and non-dusty subsamples. To further ensure the quiescent nature of the spectroscopic sample, the lower panel of Fig.~\ref{fig:spec_sample} shows the spread of the sample in the $M_{\star}$-${\rm SFR}$ plane, excluding galaxies above the dashed line that marks a 0.3 dex distance from the model by \citet{Popesso2023}, which is typically the scatter of the MS. We also plot the color-selected sample in the same panel (blue points).

Overall, dusty QG candidates show, on average, higher SFR values than the non-dusty subpopulation and the color-selected sample.
We caution that these SFR values are estimated with spectral energy distribution (SED) fitting \citepalias{Donevski2023}, which associates far-infrared (FIR) emission with obscured star formation. However, old stellar populations distributed in the diffuse ISM could also partly contribute to dust heating \citep{DaCunha2008}. Therefore, SFRs presented here can be interpreted as upper limits. We refer to the population as "quiescent" since it satisfies quiescence criteria, such as lack of optical emission lines and suitable $D_n$4000 values.

We added information on the environment by cross-matching our catalog with the density field constructed by \citet{Darvish2017}. They use a mass-complete photometric sample to compute a gridded density field on the COSMOS footprint (see \citealt{Darvish2017} for more details). Specifically, they define an overdensity value $\delta$ as the ratio between the density value of each galaxy and the median density value of the field. This allowed us to separate higher-density from lower-density environments and perform our stacking analysis in different bins.

We also retrieved information about the environment (in particular, the overdensity parameter $\delta$=$\rho$/$\bar{\rho}$) by cross-matching our catalog with the density field constructed by \citet{Darvish2017}. Finally, we assigned a morphology classification (spiral, spheroid, or irregular) to each galaxy through visual inspection of F814W HST/ACS cutouts \citep{Koekemoer2007, Massey2010}. These reveal a rich morphological diversity, including disks, bars, prominent bulges, spheroids, and irregular shapes.

\section{HI stacking results for quiescent galaxies}
\label{sec:results}
We applied our stacking pipeline (\citealt{Sinigaglia2022, Sinigaglia2024} and \citetalias{Bianchetti2025}) to the MIGHTEE-HI DR1 data, using both samples described in Sect.~\ref{sec:sample_selection} as input (see App.~\ref{sec:appendix_stacking} for the stacked spectra). Stacking QGs in the local Universe could also gain a direct comparison. MIGHTEE DR1 covers the $z<0.08$ range, but the statistics of the sample is too modest for stacking at this redshift.
For a fair comparison with the results obtained with the MIGHTEE Early Science data \citepalias{Bianchetti2025}, we derived the $M_{\star}-M_{\HI{}}$ relation for SFGs using the DR1 data and tested that it was still consistent with the previous data release (App.~\ref{sec:appendix_DR1-ES}). We divided the sample in various ways, summarized by Tab.~\ref{tab:passive_results} and discussed later in the paper.

\begin{table*}
    \caption{\label{tab:passive_results} Results from stacking runs on different QGs subsamples.}
    \centering
    \begin{tabular}{ccccccccc}
    \toprule
    \toprule
    Sample & $M_{\star}$ & IR & $D_n$4000 & Overdensity $\delta$  & Morph.Class & $N_{gal}$ & $M_{HI}$ & SNR\\
     & $[\times 10^9\, {\rm M_\odot}]$ & & & & & & $[\times 10^9\, {\rm M_\odot}]$ & \\
    \midrule
    \citepalias{Bianchetti2025} &  $8.0<\log{M_{\star}}<10.5$ & \textbackslash & \textbackslash & \textbackslash & \textbackslash & 567 & 1.70$^*$ & \textbackslash \\
    \citepalias{Bianchetti2025} &$10.5<\log{M_{\star}}<11$ & \textbackslash & \textbackslash & \textbackslash & \textbackslash & 518 & 4.58$\pm$0.58 & 8.0 \\
    \citepalias{Bianchetti2025} & $\log{M_{\star}}>11$ & \textbackslash & \textbackslash & \textbackslash & \textbackslash & 332 & 5.22$\pm$0.91 & 5.8 \\
    \midrule
    \midrule
    \citepalias{Donevski2023} &  $\log{M_{\star}}>8.0$ & no & $\forall$ & $\forall$ & $\forall$ & 118 & 3.50$^*$ & \textbackslash  \\
    \citepalias{Donevski2023} &  $8.0<\log{M_{\star}}<10.5$ & yes & $\forall$ & $\forall$ & $\forall$ & 56 & 5.26$\pm$1.31 & 4.0  \\
    \citepalias{Donevski2023} &  $10.5<\log{M_{\star}}<11$ & yes & $\forall$ & $\forall$ & $\forall$ & 119 & 5.94$\pm$1.04 & 5.8 \\
    \citepalias{Donevski2023} &  $\log{M_{\star}}>11$ & yes & $\forall$ & $\forall$ & $\forall$ & 98 & 8.15$\pm$1.19 & 6.9 \\
    \citepalias{Donevski2023} &  $\log{M_{\star}}>8.0$ & yes & $1.5<D_n4000<1.61$  & $\forall$ & $\forall$ & 79 & 6.64$\pm$1.00 & 6.7 \\
    \citepalias{Donevski2023} &  $\log{M_{\star}}>8.0$ & yes & $1.61<D_n4000<1.8$ & $\forall$ & $\forall$ & 108 & 7.76$\pm$0.85 & 9.2 \\
    \citepalias{Donevski2023} &  $\log{M_{\star}}>8.0$ & yes & $D_n4000>1.8$ & $\forall$ & $\forall$ & 98 & 4.24$\pm$1.02 & 4.2 \\
    \citepalias{Donevski2023} &  $\log{M_{\star}}>8.0$ & yes & $\forall$ & $\forall$ & Spir.\&Irr. & 132 & 7.12$\pm$0.84 & 8.5 \\
    \citepalias{Donevski2023} &  $\log{M_{\star}}>8.0$ & yes & $\forall$ & $\forall$ & Sph. & 141 & 6.04$\pm$0.95 & 6.4 \\
    \citepalias{Donevski2023} &  $\log{M_{\star}}>8.0$ & yes & $\forall$ & $\delta<1$ & $\forall$ & 109 & 7.65$\pm$1.06 & 7.2 \\
    \citepalias{Donevski2023} &  $\log{M_{\star}}>8.0$ & yes & $\forall$ & $\delta>1$ & $\forall$ & 164 & 5.86$\pm$0.77 & 7.7 \\
    \citepalias{Donevski2023} &  $\log{M_{\star}}>8.0$ & yes & $\forall$ & $\delta>1$ & Spir.\&Irr. & 75 & 6.68$\pm$0.92 & 7.3 \\
    \citepalias{Donevski2023} &  $\log{M_{\star}}>8.0$ & yes & $\forall$ & $\delta<1$ & Spir.\&Irr. & 57 & 7.71$\pm$1.48 & 5.2 \\
    \citepalias{Donevski2023} &  $\log{M_{\star}}>8.0$ & yes & $\forall$ & $\delta>1$ & Sph. & 89 & 5.13$\pm$1.29 & 4.0 \\
    \citepalias{Donevski2023} &  $\log{M_{\star}}>8.0$ & yes & $\forall$ & $\delta<1$ & Sph. & 52 & 7.58$\pm$1.75 & 4.3 \\
    \bottomrule
    \bottomrule
    \end{tabular}
    \vspace{10pt}
    \tablefoot{The first column ("Sample") lists the reference paper for the parent sample: \citepalias{Bianchetti2025} for the color-selected sample, \citepalias{Donevski2023} for the $D_n$4000-selected sample. The second column ($M_{\star}$) indicates the stellar mass range of the subsample. Columns 3 to 6 only apply to the $D_n$4000-selected sample: the third column ("IR") marks whether the galaxies in the sample have a detected Herschel counterpart and can thus be considered dusty, the fourth column ("$D_n$4000") indicates the $D_n$4000 bin (in this case, $\forall$ means $D_n$4000>1.5, as it is one of the selection criteria for quiescence). The fifth column ("$\delta$") indicates the overdensity value used as an environmental flag, and the sixth column ("Morph") marks the morphology class. The seventh column ("$N_{gal}$") indicates the number of galaxies composing the subsample, while the last two indicate the estimated HI mass after a 10\% confusion correction \citep{Sinigaglia2022, Bianchetti2025}, equipped with its uncertainty ("$M_{HI}$") and signal-to-noise ratio ("SNR"). If the \HI{} mass estimate comes with the symbol $^*$, then it refers to an upper limit at SNR=3. The $\forall$ symbol means that there is no selection for that particular value.}
\end{table*}

\subsection{A link between dust and \HI{}}
\begin{figure}
    \begin{center}
        \includegraphics{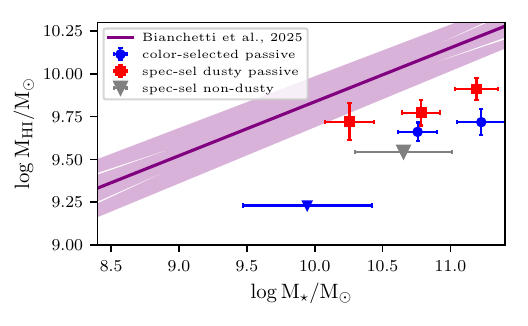}
    \end{center}
    \vspace{-0.8cm}
    \caption{Comparison of average \HI{} content between SFGs and QGs. The solid purple line represents the $M_{\rm HI}$–$M_{\star}$ scaling relation for SFGs from \citetalias{Bianchetti2025}. Blue circles show \HI{} masses from color-selected QGs in two stellar mass bins: a circle at the high-mass end (significant detection) and a triangle at the low-mass end (upper limit only). Red squares indicate \HI{} masses from stacked spectra of dusty, spectroscopically selected QGs in two mass bins. A gray triangle marks the upper limit from stacking non-dusty, spectroscopically selected QGs.}
    \label{fig:sfr_vs_ps}
\end{figure}
Fig.~\ref{fig:sfr_vs_ps} shows the results of stacking candidate QGs from the two catalogs, together with the $M_{\rm HI}$-$M_{\star}$ scaling relation for SFGs from \citetalias{Bianchetti2025}. Besides the uncertainties on the \HI{} masses (see App.~\ref{sec:appendix_stacking} for more details), we also added uncertainties on the stellar mass domain, computed as the standard deviation of the bin.
We split the QG color-selected catalog into three stellar mass bins ($8.0<\log{(M_{\star}/M_{\odot})}<10.5$, $10.5<\log{(M_{\star}/M_{\odot})}<11$ and $\log{(M_{\star}/M_{\odot})}>11$) and find no detection in the lowest mass one (blue triangle in Fig.~\ref{fig:sfr_vs_ps}), while statistically significant signals were produced in the other two stellar mass bins (blue circles). The stacked spectra are shown in App.~\ref{sec:appendix_stacking}, in the upper panels of Fig.~\ref{fig:stack1}). The \HI{} content is a factor of $\sim 3$ lower than that retained by SFGs at the same stellar mass.

For the $D_n$4000-selected sample, stacking non-dusty QGs over the full mass range only leads to an upper limit on \HI{} content (gray triangle in Fig.~\ref{fig:sfr_vs_ps}), computed as three times the RMS value obtained from the stack (Fig.~\ref{fig:stack1}, bottom panel). In contrast, dusty QGs were split into the same three stellar mass bins as for the color-selected sample (Fig.~\ref{fig:stack1}, middle panels). All three bins exhibit statistical detection of \HI{}, with the corresponding mass estimates marked by red squares in Fig.~\ref{fig:sfr_vs_ps}.

Overall, Fig.~\ref{fig:sfr_vs_ps} suggests a trend in which more massive QGs host larger \HI{} reservoirs, mirroring the $\rm M_{HI}-M{\star}$ for MS galaxies. However, QGs present diverse \HI{} content, linked to various ISM properties: color-selected QGs likely encompass a broad dust mass range, mixing dust-rich and dust-poor objects, which dilutes the stacked \HI{} signal. In contrast, the $D_n$4000-selected QGs show a correlation between dust and \HI{} content. Specifically, dust-rich galaxies possess significant \HI{} reservoirs, on average only $40\%$ less abundant than in MS galaxies. The existence of dusty candidate QGs is documented up to $z\sim3$; \citealt{Gobat2018}, \citealt{morishita22}, \citealt{Rodighiero2023}, \citealt{lee24}, \citealt{Siegel2025}). Recent studies suggest that dust-rich QGs can retain cold dust and gas for extended periods ($>2$ Gyr; \citealt{michalowski24}, \citealt{lee24},\citealt{lorenzon25}). The diversity in the dust-cold gas mass ratio \citep{Whitaker2021, morishita22, Donevski2023} suggests that dust leftovers may not be exclusively relics of star formation. Grain growth on ISM metals is proposed as a dominant dust production channel to explain the observed $M_{\rm dust}/M_{\star}$ distribution. The presence of \HI{} in dusty QGs, as reported here, is likely related to this scenario. In fact, \HI{} reservoirs may shield dust from UV radiation and favor its growth, thereby prolonging dust survival, as proposed by recent theoretical studies (\citealt{chen24}).

\subsection{Correlation with age, morphology and environment}
\begin{figure}
    \begin{center}
        \includegraphics{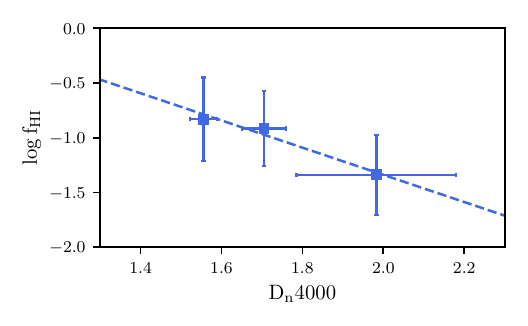}
    \end{center}
    \vspace{-0.8cm}
    \caption{Average \HI{} fraction of dusty QGs in $D_n$4000 bins. The average $f_{HI{}}=M_{\rm HI}/M_{\star}$ is shown in $D_n$4000 bins (1.5<$D_n$4000<1.61, 1.61<$D_n$4000<1.8 and $D_n$4000>1.8), plotted against their average value. The dashed line indicates the best-fit linear model in both panels.}
    \label{fig:dn4000_bins_scalrel}
\end{figure}
\begin{figure}
    \begin{center}
        \includegraphics{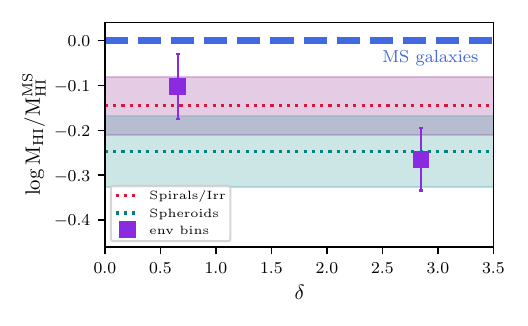}
    \end{center}
    \vspace{-0.8cm}
    \caption{Logarithm of the ratio between the average \HI{} mass of dusty QGs and the equivalent \HI{} contained at the same stellar mass in MS galaxies according to \citetalias{Bianchetti2025}, plotted against the median overdensity $\delta$ for each galaxy bin. A dashed horizontal line at y=0 indicates the \HI{} content in SFGs. Purple filled squares mark the \HI{} estimates from stacking dusty galaxies in lower-density ($\delta$<1) and in higher-density ($\delta$>1) bins, irrespective of morphology; shaded bands indicate uncertainties. Red and teal dotted lines indicate the \HI{} estimate from stacking spirals and spheroids, respectively, irrespective of environment, spanning across the $\delta$ range.}
    \label{fig:morph_env_separate}
\end{figure}
We focused on the dusty, \HI{}-rich QG sample and investigated possible dependencies on the environment and morphology. 

First, we split the sample into three $D_n4000$ bins (1.5<$D_n$4000<1.61, 1.61<$D_n$4000<1.8 and $D_n$4000>1.8) and obtained a significant detection (S/N>4) in all cases (Fig.~\ref{fig:stack2}). Fig.~\ref{fig:dn4000_bins_scalrel} shows the estimated \HI{} fractions (where $f_{\rm HI}=M_{\rm HI}/M_{\star}$) as a function of the average $D_n$4000 value in each bin. A dashed blue line indicates the best-fit linear model. Uncertainties on the $D_n$4000 axis correspond to the standard deviation within each bin. The figure suggests a trend, with older galaxies yielding lower \HI{} fractions. This implies that recently quenched systems with low $D_n$4000 have not yet consumed, expelled, or ionized their \HI{} reservoirs, or may have reaccreted \HI{} from the circumgalactic medium, through mechanisms such as tidal streams or mergers.

To investigate environmental properties, we used the overdensity parameter from \citet{Darvish2017} and defined two bins: galaxies lying in regions denser than average  ($\delta$>1) and galaxies in regions where density is lower than average ($\delta$<1). Separately stacking the two groups yields a robust signal in both cases (S/N>7, see Fig.~\ref{fig:stack3}). Fig.~\ref{fig:morph_env_separate} displays the logarithm of the ratio between the two \HI{} mass estimates in the two environment bins and the \HI{} mass expected at their average stellar mass for SFGs according to \citetalias{Bianchetti2025} (full purple squares). The higher-density stack yields an \HI{} amount $\approx$0.15 dex below the lower-density case. Environmental factors are well known to influence gas content, especially in the stellar mass range that we examine \citep{Contini2020}. \citet{Li2025} argue that \HI{} content is more strongly regulated by environment than morphology, highlighting that satellite galaxies in low-density regions often remain gas-rich. In contrast, high-density environments promote gas depletion through processes such as ram pressure stripping, tidal interactions, and starvation. Our findings are consistent with this picture, although the COSMOS field is dominated by filaments and knots rather than rich clusters, making strong environmental depletion less likely.
Recent results based on the SIMBA cosmological simulations \citep[][]{Dave2019} qualitatively support this picture \citep{lorenzon25}. Specifically, while dust- and gas-rich QGs are found in both field and cluster environments, galaxies residing in less massive halos exhibit reduced dust destruction rates. If dust and \HI{} share common removal timescales post-quenching, as observed \citep[e.g.,][]{michalowski24}, would align with our finding of slightly richer \HI{} reservoirs in lower-density environments. Moreover, the same analysis indicates that satellites may act as a secondary channel for enhancing dust and cold gas content, primarily through mergers rather than rejuvenation episodes.

Following our visual morphology classification, we stacked spirals and irregulars together, while spheroids were placed in a separate bin. We obtain clear \HI{} detections for both groups (S/N>7; see Fig.~\ref{fig:stack4}). In Fig.~\ref{fig:morph_env_separate}, a dotted crimson and teal horizontal lines mark the logarithm of the ratio between \HI{} estimates for spirals and spheroids, respectively, over the \HI{} mass for same-stellar-mass MS galaxies. The lines extend over the full $\delta$ range as they are not selected through a $\delta$ bin. The two morphology classes share a similar median $\delta$, demonstrating that they do not differ in terms of environmental properties. Dusty quiescent spirals and irregulars contain about 40\% less \HI{} mass than MS galaxies, while dusty spheroids are even more \HI{}-poor ($\approx$75\% less than MS galaxies).
\citetalias{Donevski2023} mirror these results, reporting that dusty spirals generally have a slightly higher $M_{\rm dust}$/$M_{\star}$ ratio than spheroids. 

The presence of \HI{} in quiescent spirals is not a novel result in the local Universe. \citep[e.g.,][]{Zhang2019} show that massive, quiescent disk galaxies present \HI{} fractions ranging from 10 to 30\%, comparable to those in SFGs. These studies attribute the cause of the quenching to a lack of molecular gas and dust, along with a low star formation efficiency, while \HI{} reservoirs tend to persist. The similarity in \HI{} content between our red spirals and MS galaxies suggests that these objects may be in some transitioning state --- either off the MS or toward the green valley --- i.e., experiencing a lull phase, or a post-starburst phase \citep[also see][]{ellison25}. This scenario aligns with SED fitting carried out by \citetalias{Donevski2023}, implying that the star formation histories (SFHs) of most galaxies in the sample are consistent with quenching within the past 1 Gyr. Their substantial \HI{} and dust content points to the potential for future starburst activity, placing them in a ``mini-quenching'' phase \citep{Tibor2024, Looser2025}, pending the restoration of molecular hydrogen. Moreover, a growing number of observational studies link the occurrence of \HI{}-rich QGs and post-starbursts to slow quenching mechanisms (e.g., morphological quenching), where \HI{} reservoirs persist long after the final burst (\citealt{michalowski24}, \citealt{Li2025}, \citealt{ellison25}). This conclusion is further supported by SIMBA's dust modeling, which predicts slow quenching to be prevalent among dusty QGs with old stellar populations ($>5-10$ Gyr) \citep{lorenzon25}. An independent ALMA Band 6 follow-up of 18 QGs, including several objects from this study, reveals that despite reduced molecular gas fractions, dusty QGs can retain high dust-to-gas ratios ($\gtrsim 1/100$), comparable to those of SFGs (Lorenzon et al., in preparation). They find that the total cold ISM content declines only slightly with stellar age, with quiescent spirals typically hosting somewhat richer dust reservoirs than ellipticals.

However, the fact that spheroids hold significant amounts of \HI{} --- despite their lower \HI{} content with respect to spirals --- requires some other \HI{} enrichment channel. Cutouts from the HST (App.~\ref{sec:appendix_cutouts}) reveal that spheroids are often accompanied by streams and merger remnants, suggesting that environmental properties may play a role. Although we hypothesize that high density environments favor \HI{} stripping, reaccretion may be common for quenched spheroids. Reaccretion may result from past mergers or quenching events, while frequent interaction with external gas supplies might favor accretion \citep{Dutta2019}.

\subsection{Separating morphology and environment}
\begin{figure}
    \begin{center}
        \includegraphics{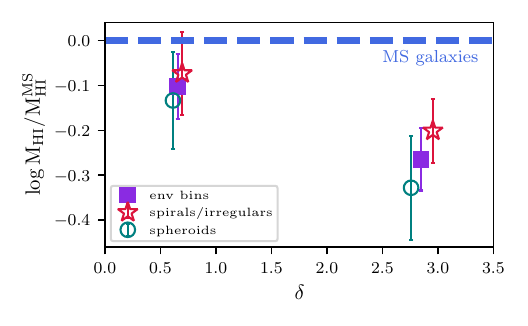}
    \end{center}
    \vspace{-0.8cm}
    \caption{Logarithm of the ratio between average \HI{} mass of dusty QGs and equivalent \HI{} mass in MS galaxies at the same stellar mas according to \citetalias{Bianchetti2025}, plotted against the median overdensity $\delta$ for each galaxy bin. A dashed horizontal line at y=0 indicates the \HI{} content of SFGs. Filled purple squares mark the \HI{} estimates from stacking dusty galaxies in lower-density ($\delta$<1) and in higher-density ($\delta$>1) bins. Empty red stars represent spiral and irregular galaxies divided into the two separate $\delta$ bins. Empty teal circles indicate spheroid galaxies in the same $\delta$ bins.}
    \label{fig:morph_env}
\end{figure}
To better unravel the impact of environment and morphology on dusty QGs, we split each morphology class into $\delta$ bins and obtained significant signal in all four cases (S/N$\geq$4, Fig.~\ref{fig:stack5}). Fig.~\ref{fig:morph_env} shows the $\log{M_{HI}/M_{HI}^{MS}}-\delta$ plane, with full purple squares marking the two $\delta$ bins. Two empty crimson stars and two empty teal circles indicate the split of the $\delta$ bins, for spirals plus irregulars and spheroids, respectively. This split shows how galaxies with different morphologies respond differently to environmental properties. Spirals in high-density environments are $\approx$30\% \HI{}-richer than their low-density peers, whereas the depletion is stronger ($\approx$60\%) for spheroids. A comparison to the square purple bins, which indicate environment bins irrespective of morphology, shows that in lower-density environments spirals and spheroids have very comparable \HI{} content (within 15\%). At higher densities spheroids are more strongly affected, retaining 30\% less \HI{} than spirals. Large error bars indicate that these results should be interpreted with caution. This suggests that although both morphology classes exhibit qualitatively similar environmental trends (i.e., decreasing \HI{} with increasing density), the effect is quantitatively stronger for spheroids. Overall, we interpret the environment to be a double-edged agent: galaxies interact with the surroundings and accrete gas from some external source, to either maintain or replenish their \HI{} supplies. However, the densest environments, such as galaxy clusters, have the opposite effect, promoting quenching through processes like ram pressure stripping, X-ray heating, tidal interactions, and harassment, likely favoring the removal of \HI{} \citep[e.g.,][]{Cortese2011, Marasco2016}. Under these conditions, larger dust reservoirs may survive if grain growth is efficient, enriching the interstellar medium with larger, more resilient grains (e.g., \citealt{aoyama19}, \citealt{nanni24}). This may also explain the presence of QGs in this study that are both dust- and \HI{}-rich regardless of environment.

\section{Conclusions} 
\label{sec:conclusions}
We present the first statistical measurement of neutral atomic hydrogen in QGs at intermediate redshift ($\langle z \rangle = 0.36$) using spectral stacking of MIGHTEE-HI data. Our analysis reveals that the presence of HI in quiescent systems is not uniform, but strongly depends on dust content, stellar mass, morphology, and environment.

We provide the highest-redshift measurement of \HI{} in QGs to date, using spectral stacking of MIGHTEE-HI data. Although direct quantitative comparison with local Universe studies among QGss studies is limited by sample selection differences and the wide spread in \HI{} measurements, our results suggest that substantial \HI{} reservoirs may be more prevalent at $\rm z \approx 0.36$, particularly in dusty systems, hinting at possible evolution in gas retention or quenching processes with redshift. Our analysis shows that the presence of \HI{} in quiescent systems is not uniform but depends on several factors, such as dust content, stellar mass, morphology, and environment.

We summarize the findings of the paper as follows:
\begin{itemize}
    \item Dust-rich QGs retain substantial \HI{} reservoirs, on average only $\approx$40\% lower than star-forming galaxies of the same mass. In contrast, dust-poor QGs show no detectable \HI{}. This highlights a close physical link between dust survival and the persistence of atomic gas.
    \item Within dusty QGs, younger systems (low $D_n$4000) are \HI{}-richer, suggesting either incomplete depletion or possible reaccretion of gas after quenching.
    \item Measured \HI{} reservoirs in spiral QGs place them systematically closer to the typical \HI{} content of star-forming galaxies. Spheroids also retain significant \HI{}, likely aided by environmental accretion or mergers. Environmental density regulates gas retention, with spheroids exhibiting greater depletion in high-density regions, while spirals are more resilient.
\end{itemize}
The detection of long-lived \HI{} reservoirs in QGs supports scenarios of slow or partial quenching, in which cold gas survives for several Gyr after star formation declines. Our results further suggest that some dusty QGs may represent transitional phases (e.g., post-starbursts or “mini-quenched” systems) with the potential for rejuvenation if molecular gas reforms.
Our findings support a scenario in which QGs do not just constitute the outcome of fuel exhaustion only: rather, a combination of internal ISM processes, morphology, and environment favors slow quenching mechanisms and enables the persistence of cold dust and gas for extended periods after quenching ($>1-2$ Gyr).
Given the large uncertainties in on our \HI{} estimates, further accuracy are needed. The current sample size may be insufficient for definitive conclusions on which parameter regulate \HI{} content in QGs. Future SKA-level sensitivities \citep{Dutta2022} will strengthen this pioneering study by providing significantly larger samples of \HI{} QGs at this redshift, possibly with statistically significant populations of direct detections.

\begin{acknowledgements}
A.B. acknowledges support of the doctoral grant funded by MIUR. 
A.B., F.S., G.R., L.B and I.P. acknowledge support from INAF under the Large Grant 2022 funding scheme (project "MeerKAT and LOFAR Team up: a Unique Radio Window on Galaxy/AGN co-Evolution”). I.P. and A.B. acknowledge financial support from the South African Department of Science and Innovation's National Research Foundation under the ISARP RADIOMAP Joint Research Scheme (DSI-NRF Grant Number 150551), and from the Italian Ministry of Foreign Affairs and International Cooperation under the “Progetti di Grande Rilevanza” scheme (project RADIOMAP, grant number ZA23GR03).
D.D. acknowledges support from the NCN through the SONATA grant UMO2020/39/D/ST9/00720. D. D thanks the support from the Polish National Agency for Academic Exchange (Bekker grant BPN /BEK/2024/1/00029/DEC/1).
M.V. and M.G. acknowledge financial support from the Inter-University Institute for Data Intensive Astronomy (IDIA), a partnership of the University of Cape Town, the University of Pretoria and the University of the Western Cape, and from the South African Department of Science and Innovation's National Research Foundation under the ISARP RADIOSKY2020 and RADIOMAP Joint Research Schemes (DSI-NRF Grant Numbers 113121 and 150551) and the SRUG HIPPO Projects (DSI-NRF Grant Numbers 121291 and SRUG22031677).
We acknowledge the use of the ilifu cloud computing facility – www.ilifu.ac.za, a partnership between the University of Cape Town, the University of the Western Cape, Stellenbosch University, Sol Plaatje University and the Cape Peninsula University of Technology. The ilifu facility is supported by contributions from the Inter-University Institute for Data Intensive Astronomy (IDIA – a partnership between the University of Cape Town, the University of Pretoria and the University of the Western Cape), the Computational Biology division at UCT and the Data Intensive Research Initiative of South Africa (DIRISA).
The MeerKAT telescope is operated by the South African Radio Astronomy Observatory, which is a facility of the National Research Foundation, an agency of the Department of Science and Innovation.
M.G. is supported by the UK STFC Grant ST/Y001117/1. For the purpose of open access, M.G. has applied a Creative Commons Attribution (CC BY) licence to any Author Accepted Manuscript version arising from this submission.
FMM carried out part of the research activities described in this paper with contribution of the Next Generation EU funds within the National Recovery and Resilience Plan (PNRR), Mission 4 - Education and Research, Component 2 - From Research to Business (M4C2), Investment Line 3.1 - Strengthening and creation of Research Infrastructures, Project IR0000034 – “STILES - Strengthening the Italian Leadership in ELT and SKA”

\end{acknowledgements}

\vspace{-20pt}

\bibliographystyle{aa} 
\bibliography{aa56367-25}

\onecolumn
\begin{appendix}

\section{Stacking method and variations with respect to \citealt{Bianchetti2025}}
\label{sec:appendix_stacking}
We hereby summarize in short the main steps of the stacking procedure outlined in \citet{Bianchetti2025}, to which we refer for an exhaustive description. We extract spectra of galaxies by collapsing subcubes defined by an aperture equal to three times the beamsize in the sky plane. Instead of using the mean beamsize across the frequency range and adopting it as a fixed reference \citepalias{Bianchetti2025}, we now create apertures as large as three times the beamsize accounting for its variation as a function of frequency. This way, aperture adapts to the frequency channel. Subcubes extend roughly $\pm$2000$\rm km s^{-1}$ in the spectral direction and we regrid them in frequency to get a homogeneous channel width of 100$\rm km s^{-1}$. Then, we assign galaxies a weight given by 1/$\sigma^{\gamma}$, where $\sigma$ is the RMS noise computed on the wings of the individual spectrum and $\gamma$ is a power index to be selected. We pick $\gamma$=1 following \citet{Fabello2011, Sinigaglia2022, Sinigaglia2024, Bianchetti2025}. At odds with \citetalias{Bianchetti2025}, we do not weigh by primary beam, as the MIGHTEE-HI DR1 data are already primary beam-corrected. Also, since the noise profile is slightly different, the RFI mask that we use to flag channels with unusual noise peaks is slightly more relaxed than the one used for Early Science data. 
\begin{figure*}[ht!]
    \begin{center}
        \includegraphics[width=\textwidth]{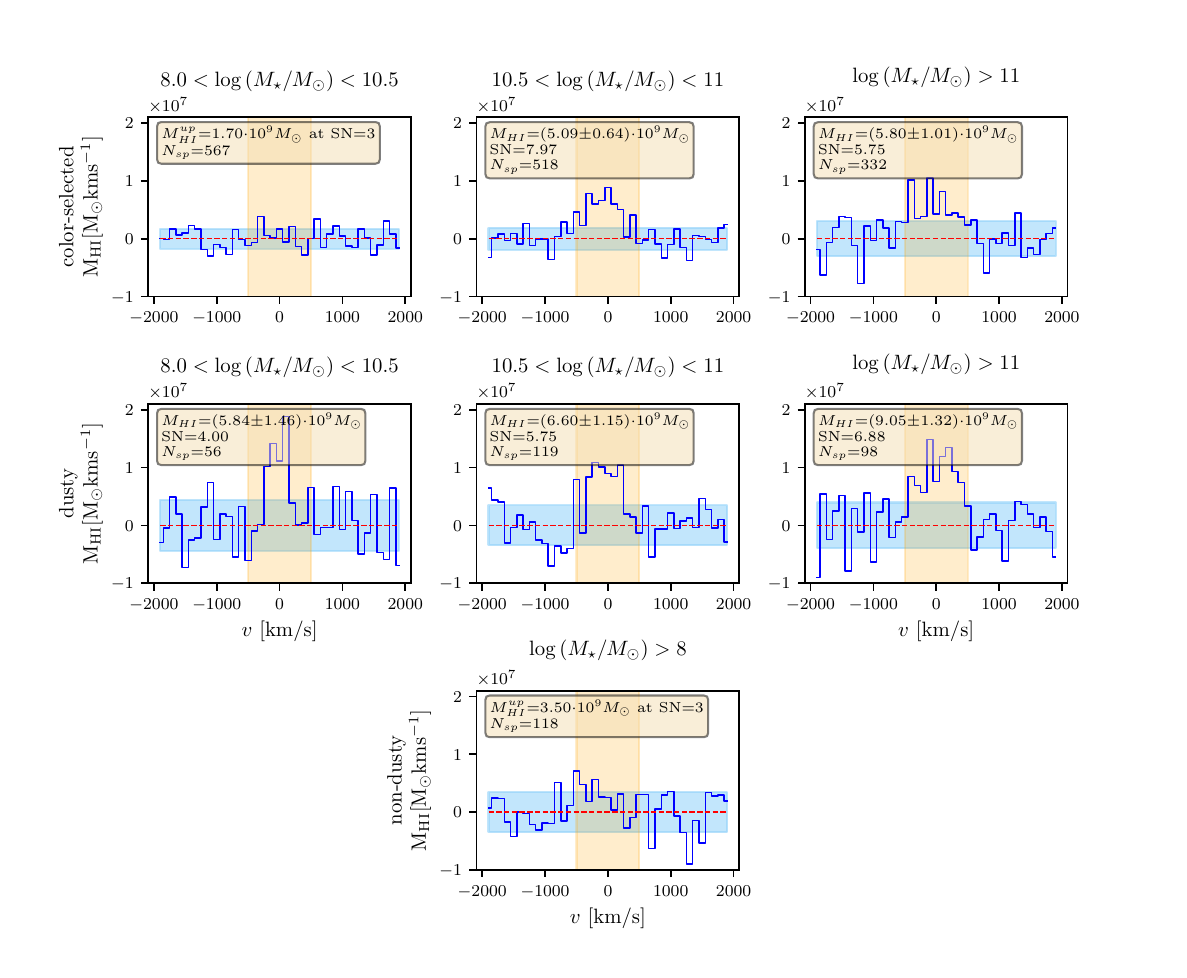}
    \end{center}
    \caption{$M_{\rm HI}$ stacks of the two different samples, photometrically selected and spectroscopically selected. All panels represent the average spectrum of the reference bin or sample. Upper panels refer to the former, split in three different stellar mass bins. Panels from the central row depict stacks from the dusty spectroscopical sample, again in the same stellar mass bins. Finally, the lower panel shows the stack of spectroscopically selected galaxies, on the whole complete range of stellar mass, with no infrared counterparts. In all cases, we use a bin width $\Delta v=100~{\rm km~s^{-1}}$. The yellow, shaded area represents the integration range in the rest-frame velocity domain, spanning from $-500~{\rm km~s^{-1}}$ to $+500~{\rm km~s^{-1}}$. The red, dotted line marks the $M_{\rm HI}$=0 line. For each panel, we report the extracted hydrogen mass and its associated uncertainty, estimated as the RMS of the bins in the region of the stacked spectrum lying outside the integration range (cyan shaded region). Finally, we report the S/N and the number of spectra selected in that stellar mass bin.}
    \label{fig:stack1}
\end{figure*}
\begin{figure*}[ht!]
    \begin{center}
        \includegraphics[width=0.9\textwidth]{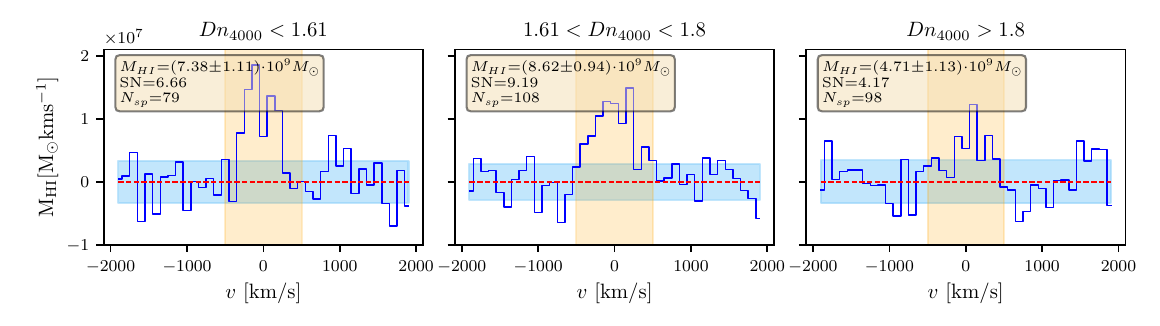}
    \end{center}
    \caption{$M_{\rm HI}$ stacks of dusty spectroscopically selected galaxies divided into three $D_n$4000 bins. Each panel represents a bin, indicated on top. See Fig.~\ref{fig:stack1} for details on color-coding.}
    \label{fig:stack2}
\end{figure*}
\begin{figure*}[ht!]
    \begin{center}
        \includegraphics[width=0.85\textwidth]{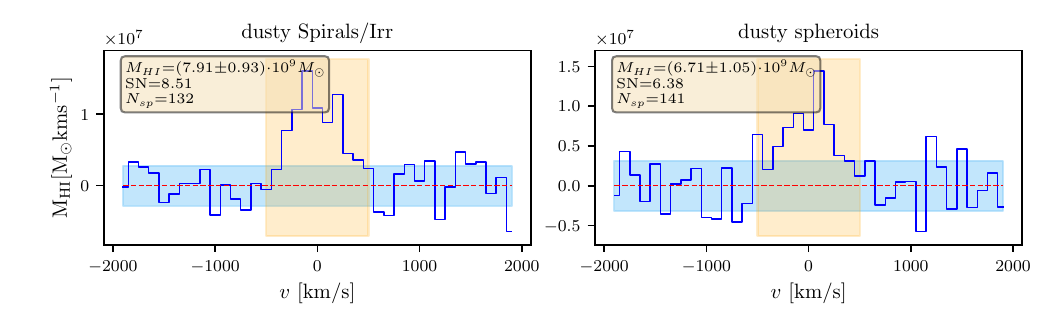}
    \end{center}
    \caption{$M_{\rm HI}$ stacks of dusty spectroscopically selected galaxies divided into two morphology bins: on the left panel, spirals and irregulars are stacked together; on the right panel, we show the stack of spheroids. See Fig.~\ref{fig:stack1} for details on color-coding.}
    \label{fig:stack3}
\end{figure*}
\begin{figure*}[ht!]
    \begin{center}
        \includegraphics[width=0.85\textwidth]{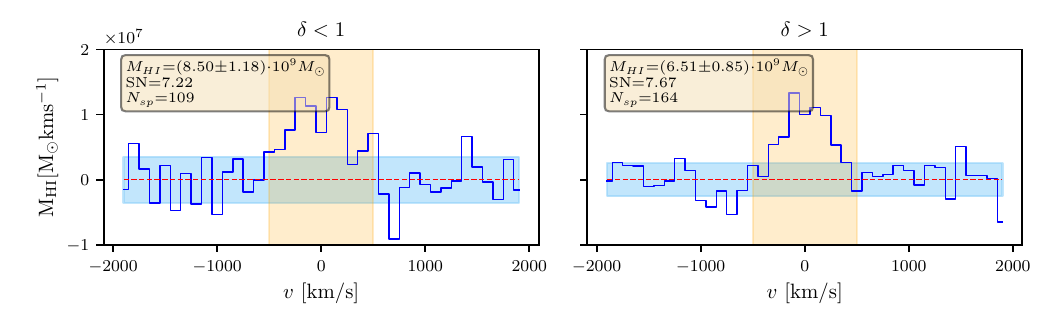}
    \end{center}
    \caption{$M_{\rm HI}$ stacks of dusty spectroscopically selected galaxies divided into two environment bins: on the left panel, we show the stack of galaxies in low-density regions, marked by overdensities $\delta<$1; the right panel displays the stack of galaxies in higher density environments. See Fig.~\ref{fig:stack1} for details on color-coding.}
    \label{fig:stack4}
\end{figure*}
\begin{figure*}[ht!]
    \begin{center}
        \includegraphics[width=0.85\textwidth]{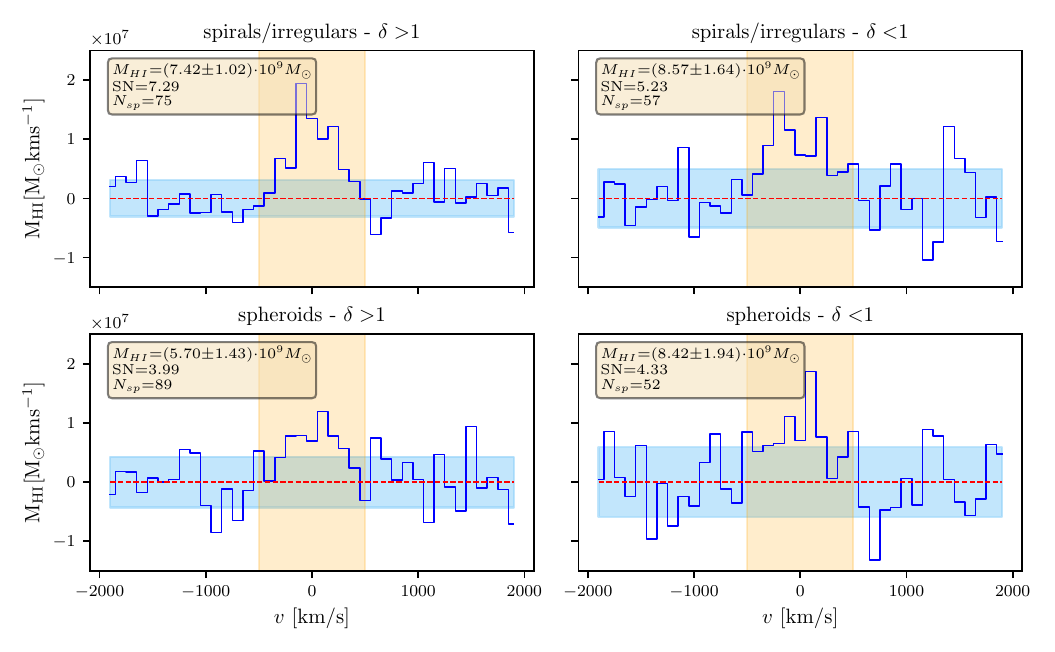}
    \end{center}
    \caption{$M_{\rm HI}$ stacks of dusty spectroscopically selected galaxies divided into four different groups, obtained by crossing morphology and environment phase space. Morphology class and environment selection condition is indicated on top each panel. See Fig.~\ref{fig:stack1} for details on color-coding.}
    \label{fig:stack5}
\end{figure*}

\HI{} mass is computed by integrating the final, stacked spectrum within a $\pm$500 $\rm km s^{-1}$ spectral range. This is slightly larger than what was done in \citetalias{Bianchetti2025}, since stacked spectra in this work are generally broader. The reason for that might be linked to an intrinsically higher accuracy of the spec-z of SFGs, whose spectrum is typically crowded with emission lines, which makes redshift determination more precise. Uncertainties on \HI{} mass are given by the RMS computed on the wings of the spectrum, i.e. the channels lying outside the integration region.

Figures ~\ref{fig:stack1}-~\ref{fig:stack5} show the stacks from which we estimate the \HI{} masses used in the analysis. Note that ~\ref{fig:stack1} sustains the claim that non-dusty galaxies contain no \HI{}. We have tested whether or not the non-detection of \HI{} in non-dusty galaxies solely depends on the limited size of the sample: in fact, non-dusty galaxies in the sample are roughly as numerous as one-third of the total dusty galaxies. However, the stack of several realizations of a subset of the dusty QGs, as big as the sample of non-dusty, always yielded a detection, proving that this is a physically driven result. Thus, we can assume that no \HI{} above the depicted upper limit (gray triangle in Fig.~\ref{fig:sfr_vs_ps}) is contained in non-dusty galaxies.

\section{Consistency with scaling relation}
\label{sec:appendix_DR1-ES}
In this work we stack candidate QGs from a spectroscopic catalog on the MIGHTEE-HI DR1 data \citep{Heywood2024}: among the several data products available, we use the \HI{} datacubes reduced with a robustness parameter $r=0.0$, with a final synthesized beam around 20" and spectral resolution of 105 KHz. and with no additional polynomial image-plane continuum subtraction, as we perform our own baseline correction as explained in Sect.~\ref{sec:appendix_stacking}. Throughout the paper, we make direct comparison with the $M_{\rm HI}$-$M_{\star}$ scaling relation extracted by \citetalias{Bianchetti2025} from the MIGHTEE-HI Early Science Data (ES) \citep{Maddox2021}.
\begin{figure*}[ht!]
    \begin{center}
        \includegraphics{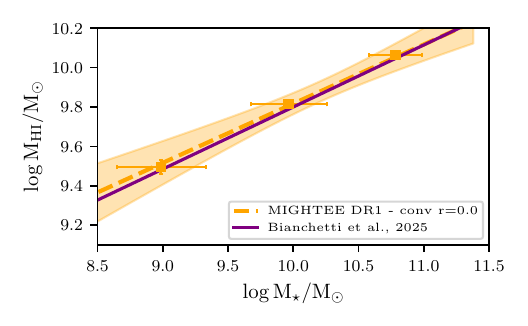}
    \end{center}
    \caption{Comparison between the $M_{\rm HI}$-$M_{\star}$ scaling relation from \citetalias{Bianchetti2025} (purple solid line) and the $M_{\rm HI}$ values extracted from the stacks performed on the MIGHTEE-DR1 data with the same settings (orange squares), fitted by a linear model (dashed orange line).}
    \label{fig:DR1-ES}
\end{figure*}
To make sure these results can be consistently compared, considering both the change in dataset and the slight modifications of the pipeline discussed in Appendix ~\ref{sec:appendix_stacking}, we stack MIGHTEE DR1 spectra extracted the same catalog of SFGs that was used in \citetalias{Bianchetti2025}. Results of this consistency check are illustrated in Fig.~\ref{fig:DR1-ES}, where the solid purple line marks the scaling relation reported by \citetalias{Bianchetti2025}, while the orange dots represent the \HI{} masses inferred from the stacks performed on the three MIGHTEE mass bins (8.0<$\log{M_{\star}}$<9.5, 9.5<$\log{(M_{\star}/M_{\odot})}$<10.5, $\log{(M_{\star}/M_{\odot})}$>10.5) described in that paper. All points are consistent with the solid line within the uncertainties, proving that the ES and DR1 datasets are completely consistent, as already demonstrated by \citet{Heywood2024}.

\section{Morphology classification}
\label{sec:appendix_cutouts}
\begin{figure*}[ht!]
    \begin{center}
        \includegraphics[width=.6\textwidth]{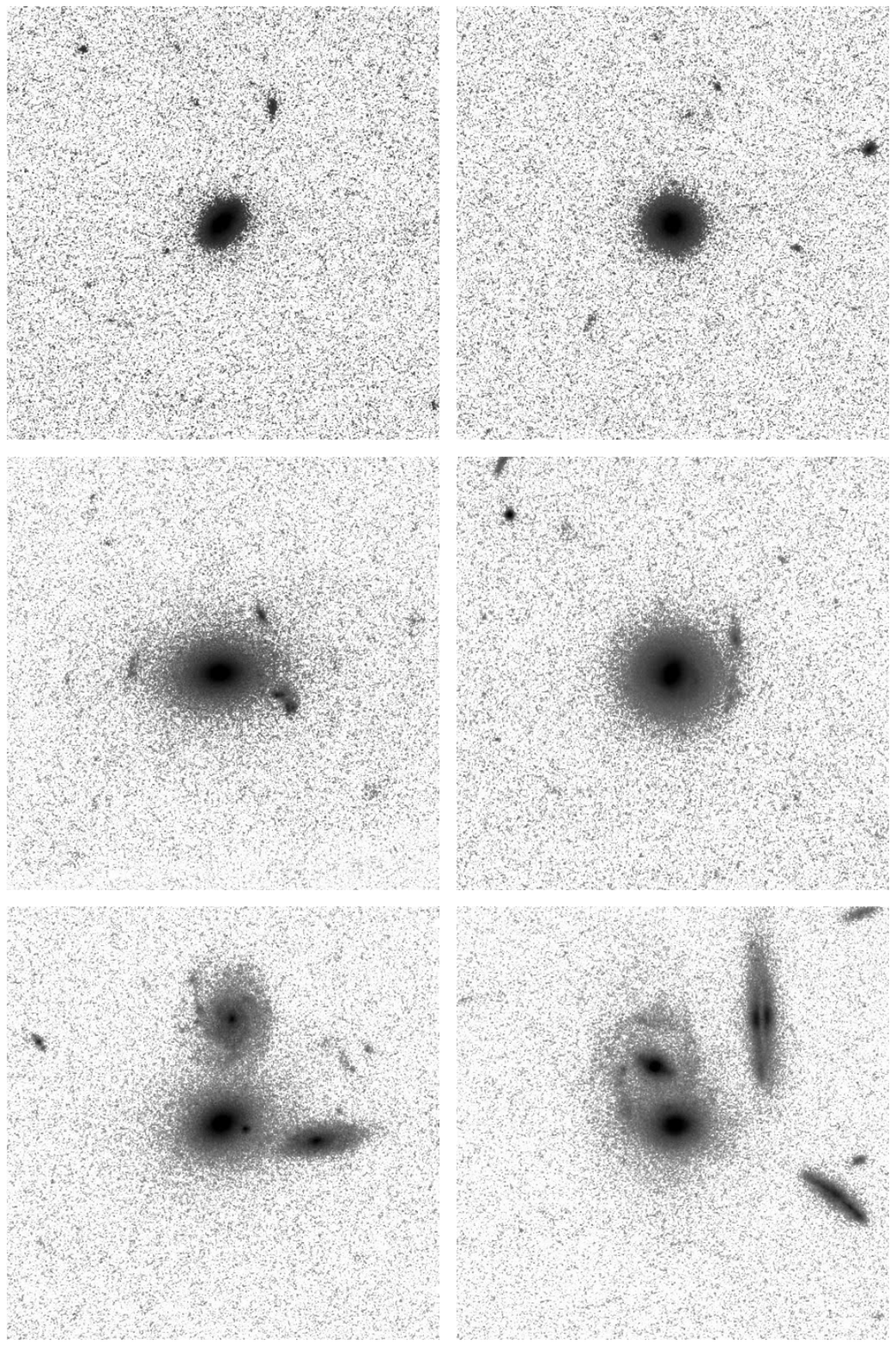}
    \end{center}
    \caption{15"x15" ACS/HST cutouts for dusty spheroids. Cases for isolated systems (top row), accretion through minor mergers (middle row) and spheroids in groups (bottom row) are shown.}
    \label{fig:cutouts}
\end{figure*}

In this appendix we provide some example of the visual classification that we performed on the sample of dusty QGs drawn from \citepalias{Donevski2023}.
We checked ACS/HST images over the COSMOS field in the F814W filter. Sources are separated into spirals, irregulars and spheroids. The latter class includes a variety of environmental signatures, that might be connected to the observed enhancement of dust and HI in this specific sample. Figure ~\ref{fig:cutouts} illustrates cutouts of size 15"x15" (corresponding to $\sim80$ kpc at $z\sim0.37$) for: isolated spheroids (top row), spheroids with potential satellites or ongoing minor mergers (middle row), spheroids in groups (bottom row). 

\end{appendix}

\end{document}